# HART: A Hybrid Addressing Scheme for Self-Balancing Binary Search Trees in Phase Change Memory (PCM)

1st Mahek Desai
Computer Science Department
California State University, Northridge
Northridge, USA
mahek.desai.849@my.csun.edu

2nd Apoorva Rumale
Computer Science Department
California State University, Northridge
Northridge, USA
apoorva-sanjay.rumale.462@my.csun.edu

3rd Marjan Asadinia
Computer Science Department
California State University, Northridge
Northridge, USA
marjan.asadinia@csun.edu

*Abstract*—As DRAM and other transistor-based memory technologies approach their scalability limits, alternative storage solutions like Phase-Change Memory (PCM) are gaining attention for their scalability, fast access times, and zero leakage power. However, current memory-intensive algorithms, especially those used in big data systems, often overlook PCM's endurance limitations (($10^6$ to $10^8$ writes before degradation) and write asymmetry. Self-balancing binary search trees (BSTs), which are widely used for large-scale data management, were developed without considering PCM's unique properties, leading to potential performance degradation. This paper introduces HART, a novel hybrid addressing scheme for self-balancing BSTs, designed to optimize PCM's characteristics. By combining DFAT-Gray code addressing for deeper nodes with linear addressing for shallower nodes, HART balances reduced bit flips during frequent rotations at deeper levels with computational simplicity at shallow levels. Experimental results on PCM-aware AVL trees demonstrate significant improvements in performance, with a reduction in bit flips leading to enhanced endurance, increased lifetime, and lower write energy and latency. Notably, these benefits are achieved without imposing substantial computational overhead, making HART an efficient solution for big data applications.

## I. INTRODUCTION

The exponential growth in data-driven applications has marked the beginning of the big data era, creating unprecedented demands for efficient and scalable memory systems. [7]. Phase-Change Memory (PCM) has emerged as a promising alternative to DRAM and other transistor-based memory technologies, addressing critical challenges such as scalability limitations, high leakage power, and energy inefficiencies [11], [5]. PCM's attributes, including high density, non-volatility, and byte addressability [8], [14], make it a strong candidate for next-generation computing systems, especially in the context of big data applications. However, PCM is not without its challenges, including limited endurance, write asymmetry, and latency concerns, which are exacerbated by write-intensive workloads [9], [10], [12]. Among data structures, self-balancing binary search trees, such as AVL trees, play a pivotal role in efficiently managing and accessing large-scale data, a cornerstone of big data analytics [13]. However, traditional implementations of these trees are not optimized for PCM, often leading to excessive write operations and subsequent degradation of PCM cells. These challenges necessitate innovative solutions that can leverage the strengths of PCM while addressing its inherent limitations [2]–[4], [10], [12].

This paper presents HART, a novel Hybrid Addressing scheme for self-balancing binary search trees, tailored specifically for PCM-based systems. The proposed scheme synergistically combines DFAT-Gray code addressing for deeper tree nodes with linear addressing for shallower nodes. This hybrid approach achieves a critical balance: minimizing bit flips during frequent rotations in deeper levels while maintaining computational simplicity at shallower levels. By reducing write energy and latency, HART significantly enhances PCM's endurance and operational efficiency.

Experimental evaluations demonstrate that HART outperforms existing addressing schemes in reducing bit flips and computational cost, providing a scalable and efficient solution for PCM-aware data structures in big data environments. This contribution not only addresses the technical challenges of PCM but also aligns with the broader goals of enhancing memory architecture to meet the demands of modern data-intensive applications.

## II. MOTIVATION

To manage and access massive datasets efficiently, self-balancing binary search trees are commonly employed due to their capability to limit access latency while maintaining optimal space utilization. One such tree is the AVL tree, a widely used data structure that ensures balanced height through rotations during insertions and deletions. This balancing mechanism minimizes the worst-case access latency, providing logarithmic time complexity for search, insertion, and deletion operations.

DFAT-Gray [1] addresses the challenge of write efficiency in data structures for Phase-Change Memory (PCM), specifically due to its write asymmetry and endurance limitations. The authors redesign self-balancing binary search trees, such as AVL trees, to minimize bit flips during rotations—one of the main overheads in PCM-based systems. Using a Depth-First Alternating Traversal (DFAT) approach for grouping nodes, which traverses the tree while alternating directions between left and right subtrees, ensures that related nodes are assigned indices with minimal Hamming distance. This reduces the number of bit flips during AVL tree rebalancing. Once the index is calculated, the Binary-to-Gray Conversion algorithm is applied, converting binary indices to Gray code addresses using a simple XOR operation between the bits. This conversion guarantees addresses with minimal Hamming distance are assigned to adjacent nodes, further reducing the impact of write operations on PCM cells. This study focuses on static allocation; however, our work extends these concepts by introducing a hybrid allocation scheme. This extension optimizes both endurance and write efficiency, filling a critical gap in adaptive memory management strategies for multi-type systems.

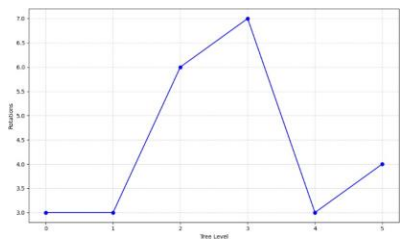
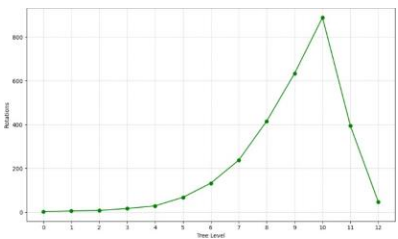
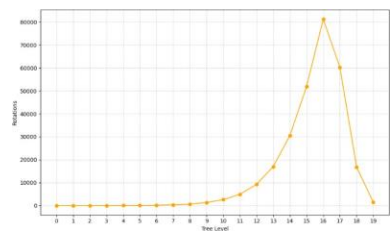

Fig. 1: Rotations per Level for 8 bits per pointer.

Fig. 2: Rotations per Level for 15 bits per pointer.

Fig. 3: Rotations per Level for 22 bits per pointer.

### A. Observation on Tree Rotations

The frequency and distribution of rotations across different tree levels directly impact the potential for bit flip reduction, as each rotation operation involves pointer modifications that could trigger multiple bit transitions. To validate our hypothesis that deeper tree levels experience a higher frequency of rotations, we conducted a comprehensive analysis using various bit-width configurations for pointers.

We constructed tree rotations for configurations ranging from 8 bits to 22 bits per pointer to analyze trends and their implications systematically. Specifically, we plotted three key graphs representing two extreme cases and an intermediate configuration: 8 bits (minimal), 15 bits (moderate), and 22 bits (extensive) per pointer. These configurations were chosen to ensure a comprehensive understanding of rotation patterns across varying tree capacities and depths. The experimental results, as illustrated in Figures 1, 2, and 3, highlight rotation frequency distributions that reveal compelling patterns consistent with our initial hypothesis. Notably, this trend was observed across all bits-per-pointer configurations, further substantiating the robustness of our findings.

In AVL trees, the rotation frequency at deeper levels increases exponentially due to several factors. First, rotations at deeper levels often trigger a cascading effect, requiring additional balancing operations up the tree. Second, deeper levels exhibit increased sensitivity to imbalances as the height differences between subtrees become more critical for maintaining the AVL property. Third, operations affecting deeper nodes involve traversing longer paths from the root, multiplying the potential rotation points. As a result, there is a significant surge in rotation frequency at deeper levels, with frequencies peaking at around 7 at level 3 for the minimal configuration, approximately 900 at level 10 for the moderate configuration, and 80,000 at level 15 for the extensive configuration. This pattern underscores the importance of optimizing bit flip reduction strategies for deeper tree levels in DFAT-Gray encoding to improve energy efficiency and minimize computational overhead.

### III. METHODOLOGY

In this section, we describe the core components of the Hybrid Addressing Scheme proposed for PCM-aware AVL trees, focusing on the algorithms responsible for efficient address assignment based on depth and tree rotations. The method integrates Depth-First Alternating Traversal (DFAT) along with gray code addressing for deeper levels and linear addressing for shallow levels, effectively balancing computational cost and bit flip reduction. The explanation below breaks down the key algorithms involved, including the Hybrid Address Assignment, Threshold Selection and Comparison. Figures 4, 5 and 6 present the visual representation of HART across varying threshold values. Details about threshold calculation will be discussed briefly later in the paper in section III-B1.

### A. Overview

This section entails an in-depth discussion of our proposed method - HART, a Hybrid Addressing Scheme that combines DFAT-Gray addressing with linear addressing. This hybrid strategy leverages the benefits of both approaches, optimizing the trade-off between computational efficiency and bit flip reduction. By dynamically switching addressing schemes based on a threshold $T$, derived as a fraction of the tree height $H$, the proposed method ensures a more practical and scalable implementation for PCM-aware AVL trees. In this scheme:

- **Shallower levels** (depth $\leq T$): Linear addressing is employed, where rotations are rare, and computational simplicity takes precedence.

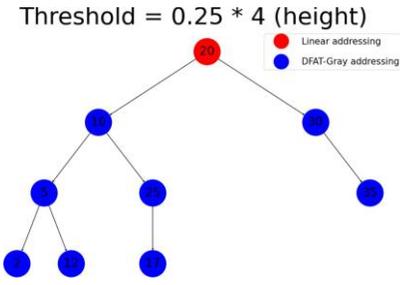

Fig. 4: Visual Representation of HART at 0.25H Threshold.

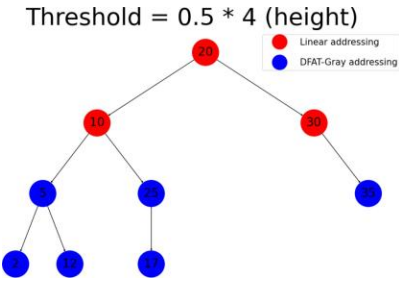

Fig. 5: Visual Representation of HART at 0.5H Threshold.

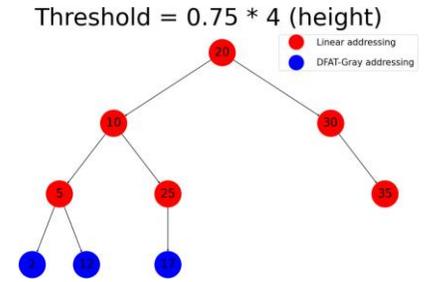

Fig. 6: Visual Representation of HART at 0.75H Threshold.

- **Deeper levels** (depth $> T$): DFAT-Gray addressing is applied, as these levels experience frequent rotations, making bit flip minimization critical.

### B. Hybrid Addressing Scheme for PCM-Aware AVL Trees

*1) Threshold Selection and Comparison (Algorithm 1):* The Threshold Selection and Comparison algorithm plays a crucial role in evaluating the efficiency of the proposed hybrid addressing scheme by comparing the performance at three different threshold ratios. These ratios, defined as $\frac{1}{4}, \frac{1}{2}, \frac{3}{4}$ or 0.25, 0.5, and 0.75 respectively, of the tree height, allow for a systematic analysis of how varying the threshold impacts performance in terms of bit flips and computational cost.

*While numerous thresholds were considered during the study, we have presented results for $\frac{1}{4}, \frac{1}{2}, \frac{3}{4}$ because they not only represent evenly distributed fractions of the tree height, enabling a balanced evaluation of thresholds across the lower, middle, and upper ranges of the tree but also adequately demonstrate the consistent trend observed across the range of varying thresholds. Moreover, the selection of an appropriate threshold can be tailored to specific case needs and requirements, ensuring alignment with the intended use case. As shown in Section IV, we observed that regardless of the chosen threshold, the overall trend remains consistent: as the threshold increases, the number of bit flips rises, while the computational cost decreases.*

- **Threshold Calculation:** For each threshold ratio, the algorithm calculates the threshold $T$ based on the height $H$ of the tree, which is determined by equations 1 and 2.

$$H = \lceil \log_2(\text{num\_nodes} + 1) \rceil \tag{1}$$

$$T = \text{Ratio} \cdot H \tag{2}$$

The threshold is then set by multiplying the tree height by the ratio and rounding to the nearest integer. This threshold is used to determine which nodes will use DFAT-Gray addressing and which will use linear addressing.

- **Performance Metrics:** The algorithm runs the hybrid addressing scheme for each threshold ratio and collects performance metrics, specifically the number of bit flips and computational cost for each configuration. These results provide insights into the trade-offs between bit flip reduction and computational cost for different thresholds.
- **Comparison:** The algorithm returns the comparison results, highlighting the impact of the threshold selection on the performance of the hybrid addressing scheme.
- **Complexity:** The algorithm iterates over a fixed set of ratios and processes $n$ nodes for each threshold evaluation, leading to an overall complexity of $O(n)$.

By systematically testing different threshold values, this algorithm enables the selection of the most efficient configuration for the Hybrid Addressing Scheme, ensuring the best balance between minimizing bit flips and maintaining fast computational performance.

---

**Algorithm 1** Threshold Selection and Comparison

1: **Input:**
2:   *num_nodes*: The total number of nodes in the tree.
3: **Output:**
4:   A comparison of bit flips and computational cost for different threshold values.
5: **procedure** COMPARETHRESHOLDS(*num_nodes*)
6:   Initialize a list of threshold ratios: *threshold_ratios* ← {1/4, 1/2, 3/4}
7:   Initialize an empty results dictionary: *results* ← {}
8:   **for** each *ratio* in *threshold_ratios* **do**
9:     Calculate the maximum depth of the tree: *depth* ← ⌈log₂(*num_nodes* + 1)⌉
10:    Calculate the threshold for the current ratio: *threshold* ← max(1, round(*depth* × *ratio*))
11:    Run the Hybrid Addressing Scheme with the calculated threshold.
12:    Collect bit flip and time metrics for the current configuration.
13:    Add the collected metrics to the *results* dictionary.
14:  **end for**
15:  Return the *results* dictionary.
16: **end procedure**

---

*2) Hybrid Address Assignment (Algorithm 2):* The primary objective of the Hybrid Address Assignment algorithm is to assign memory addresses to nodes in an AVL tree efficiently, depending on the depth of each node and the chosen threshold. The algorithm recursively traverses the tree, adjusting the address assignment strategy based on the node's depth relative to the threshold $T$.

- **Depth-based Addressing:** For nodes at a depth greater than the threshold $T$, DFAT-Gray addressing is applied to minimize bit flips during rotations. This ensures that frequently rotated nodes, typically located at deeper tree levels, experience minimal bit flip overhead during pointer updates.
- **Linear Addressing:** For shallower tree levels, where rotations are less frequent, the algorithm assigns linear addresses to nodes. This simplifies the computational overhead associated with address assignment, ensuring that computational cost does not grow unnecessarily as the tree depth increases.
- **Recursive Assignment:** The algorithm operates recursively, starting from the root of the AVL tree, and propagates downwards to the left and right child nodes. At each level, it assigns the address to the current node and then recursively processes its left and right children. This ensures that all nodes are assigned an address based on their depth and the threshold.
- **Complexity:** The algorithm processes $n$ nodes, and each node's address assignment involves operations like Gray code conversion ($O(\log n)$) and DFAT index calculation ($O(\log n)$). This results in an overall complexity of $O(n \log n)$.

The recursive nature of the algorithm ensures that it applies the appropriate addressing scheme (DFAT-Gray or linear) depending on the node's position in the tree, optimizing both performance and bit flip reduction. The entire process of threshold selection, comparison, and hybrid address assignment for an AVL tree with $n$ nodes has a total complexity of $O(n \log n)$.

## IV. EVALUATION

### A. Experimental Setup

This section describes the experimental setup used to evaluate the proposed Hybrid Addressing Scheme for PCM-aware AVL trees. We test the scheme using different numbers of data items and varying address space sizes. The experimental setup and the methods used are explained below and can also be seen in Table I.

TABLE I: Experimental Setup

| Parameters | Value Range |
|---|---|
| Number of Bits in a Pointer | 8 to 21 |
| Number of Nodes | $2^6 - 1$ to $2^{19} - 1$ |
| Tree Height | 6 to 19 Levels |
| Threshold Ratios | 0.25, 0.5 and 0.75 |

*1) Data Set Generation and Insertion Process:* For each experiment, an input data set with $N$ items is generated, where $N$ ranges from $2^6 - 1$ to $2^{19} - 1$. This range allows evaluating performance across various tree sizes, from small to large, to assess scalability. The data array, consisting of integers from 0 to $N - 1$, is shuffled randomly by iteratively selecting items and swapping them until fully randomized. This array is then used for AVL tree insertion.

The method's scalability makes it ideal for commercial PCM applications and real-world datasets, ensuring efficient memory management. It adapts to increasing dataset sizes with minimal computational overhead, enabling reliable deployment in large-scale PCM-based storage systems.

For an AVL tree with $H + 2$ levels, $H + 2$ bits are required to represent each node's address, where $H$ is the tree height from equation 1. Since a full binary tree with $x$ levels can have up to $2^x - 1$ nodes, the pointer bits range from 8 to 21. Each setting of $2^x - 1$ nodes, where $x$ is from 6 to 19, is tested 100 times to compute average bit flips per rotation and computational cost. Based on our observations, we are confident that the trends identified in this range will remain consistent for even larger datasets.

*2) Address Allocation Methods:* We compare five address allocation methods to evaluate their effectiveness in reducing bit flips and computational cost during node insertion. These methods are as follows:

- **HART:** Our proposed solution, where DFAT-Gray addressing is applied at deeper tree levels (where rotations are frequent) and linear addressing is applied at shallower tree levels (where rotations are less frequent). This hybrid approach aims to strike a balance between reducing bit flips and minimizing computational overhead.

---

**Algorithm 2** Hybrid Address Assignment

1: **Input:**
2:    *node*: The current node in the tree.
3:    *level*: The current depth in the tree.
4:    *hybrid_threshold*: The threshold level for switching to linear addressing.
5: **Output:**
6:    Assigns an address to each node in the tree.
7:    **procedure** ASSIGNADDRESS(*node, level, hybrid_threshold*)
8:       **if** *node* is null **then**
9:          **return**
10:       **end if**
11:       **if** *level* is less than or equal to *hybrid_threshold* **then**
12:          Assign the next available linear address to the node.
13:          Increment the next linear address counter.
14:       **else**
15:          Calculate the DFAT index for the node's key, relative to the root node.
16:          Convert the DFAT index to a Gray code and assign it as the node's address.
17:       **end if**
18:       Add the node's address to the address space.
19:       Recursively call the 'AssignAddress' procedure for the left and right children of the node, incrementing the level.
20: **end procedure**

- **DFAT-Gray**: Depth-First Alternating Traversal (DFAT) is used to index each node, and the Gray code is used to assign addresses for all nodes. The Gray code minimizes the Hamming distances of adjoining nodes, reducing bit flips during tree rotations. This method serves as the baseline for comparison.
- **Gray:** Gray code is used for converting binary numbers into a unique address system where each adjacent node has addresses with minimal Hamming distance. This is achieved through a simple XOR operation between adjacent bits in the binary representation.
- **Linear:** This method sequentially allocates addresses to nodes as they are inserted. It begins with an address of 0 and increments the value with each subsequent assignment. This technique is straightforward and comparatively less computationally intensive.
- **Random:** In this method, an available address is chosen randomly and assigned to each newly inserted node.

### B. Experimental Results

The performance of the proposed HART encoding schemes with threshold = 25%, 50%, and 75% of tree height — was evaluated against the baseline DFAT-Gray, Linear, Random and Gray addressing schemes across various metrics, including the number of bit flips and average computational cost. The experiments were conducted for pointers of varying lengths, ranging from 8 to 21 bits, corresponding to addressing spaces with nodes ranging from $2^6 - 1$ to $2^{19} - 1$.

In HART schemes, the percentage (25%, 50%, or 75%) denotes the proportion of levels whose nodes use the linear addressing, while the remaining nodes adopt DFAT-Gray code addressing scheme. For instance:

- **HART at 0.25H Threshold**: Assigns linear addressing to the top 25% of the nodes (shallow levels) and uses DFAT-Gray encoding for the remaining 75% (deeper levels).
- **HART at 0.5H Threshold**: Divides the addressing evenly, with DFAT-Gray used for the bottom 50% of the nodes and linear addressing for the top 50%.
- **HART at 0.75H Threshold**: Assigns linear addressing to the top 75% of the nodes and DFAT-Gray encoding to the bottom 25%.

These configurations aim to balance the reduced bit flips during frequent rotations in deeper tree levels with computational simplicity at shallower levels, where rotations are less common. The evaluation highlights the trade-offs achieved by each hybrid scheme, showcasing improvements in write efficiency and endurance of PCM-based AVL trees.

*1) Average Computational Cost:* The average computational cost results, depicted in Figure 7, reveal a significant improvement in performance for HART schemes compared to DFAT-Gray. While DFAT-Gray's computational cost scales exponentially with pointer length, the hybrid schemes maintain a notably lower computational cost. For instance, at 14 bits per pointer, DFAT-Gray exhibits a computational cost of 36.59 s, whereas HART at 0.75H Threshold achieves a mere 7.32 s, reflecting a substantial efficiency gain. Notably, HART at 0.25H Threshold and HART at 0.5H Threshold follow similar trends with computational cost of 18.3 s and 12.8 s, respectively, further emphasizing the trade-off between hybridization level and computational overhead. Table II presents the computational cost comparison for different addressing schemes based on the number of bits in a pointer. Gray, linear, and random addressing schemes are not included in the comparison because it is already well-established that these schemes inherently incur lower computational costs.

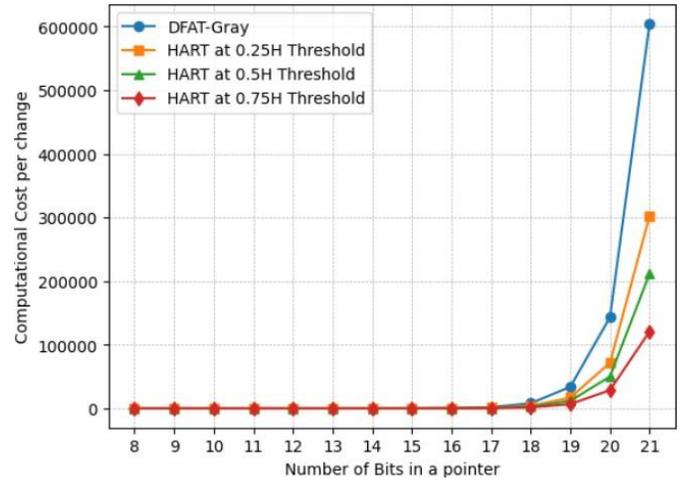

Fig. 7: Computational Cost (s) for respective number of bits in a pointer.

TABLE II: Computational Cost per Number of Bits in a Pointer

| No. of bits in Pointer | DFAT-Gray (s) | HART at T = 0.25H (s) | HART at T = 0.5H (s) | HART at T = 0.75H (s) |
| --- | --- | --- | --- | --- |
| 8  | 0.0063    | 0.00315     | 0.002205    | 0.00126    |
| 9  | 0.0283    | 0.01415     | 0.009905    | 0.00566    |
| 10 | 0.1118    | 0.0559      | 0.03913     | 0.02236    |
| 11 | 0.4526    | 0.2263      | 0.15841     | 0.09052    |
| 12 | 1.9955    | 0.99775     | 0.698425    | 0.3991     |
| 13 | 8.6897    | 4.34485     | 3.041395    | 1.73794    |
| 14 | 36.5893   | 18.29465    | 12.806255   | 7.31786    |
| 15 | 67.4775   | 33.73875    | 23.617125   | 13.4955    |
| 16 | 461.4869  | 230.74345   | 161.520415  | 92.29738   |
| 17 | 1911.963  | 955.9815    | 669.18705   | 382.3926   |
| 18 | 8035.3157 | 4017.65785  | 2812.360495 | 1607.06314 |
| 19 | 33893.3577 | 16946.67885 | 11862.6752 | 6778.6715  |
| 20 | 143095.77 | 71547.885   | 50083.5195  | 28619.154  |
| 21 | 604281.63 | 302140.815  | 211498.5705 | 120856.326 |

*2) Number of Bit Flips:* Our comprehensive analysis demonstrates that HART consistently outperforms traditional addressing schemes, with an average reduction of 84.7% in bit flips compared to random addressing, 73.2% compared to Gray coding, and 76.5% compared to linear addressing. As shown in Figure 8, the HART schemes exhibit a controlled increase in the number of bit flips compared to DFAT-Gray. The HART at 0.25H Threshold scheme results in an average increase of 11.2% in bit flips relative to DFAT-Gray. Similarly, the HART at 0.5H Threshold and HART at 0.75H Threshold schemes exhibit increases of 27.7% and 41.4%, respectively.

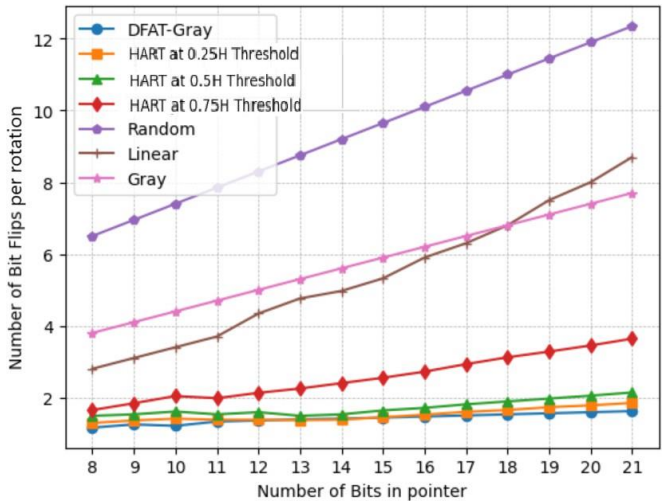

Fig. 8: Number of Bit Flips for respective bits in a pointer.

TABLE III: Bit Flip Costs per Number of Bits in a Pointer

| No. of bits in Pointer | DFAT-Gray | HART at T = 0.25H | HART at T = 0.5H | HART at T = 0.75H |
|---|---|---|---|---|
| 8 | 1.16 | 1.29 | 1.49 | 1.65 |
| 9 | 1.25 | 1.36 | 1.53 | 1.84 |
| 10 | 1.21 | 1.41 | 1.61 | 2.04 |
| 11 | 1.33 | 1.38 | 1.53 | 1.98 |
| 12 | 1.36 | 1.38 | 1.59 | 2.13 |
| 13 | 1.39 | 1.37 | 1.49 | 2.25 |
| 14 | 1.41 | 1.38 | 1.53 | 2.40 |
| 15 | 1.44 | 1.45 | 1.64 | 2.55 |
| 16 | 1.47 | 1.52 | 1.71 | 2.72 |
| 17 | 1.50 | 1.60 | 1.81 | 2.93 |
| 18 | 1.53 | 1.65 | 1.89 | 3.12 |
| 19 | 1.56 | 1.73 | 1.97 | 3.28 |
| 20 | 1.59 | 1.78 | 2.05 | 3.45 |
| 21 | 1.62 | 1.85 | 2.14 | 3.64 |

However, the trend remains consistent across pointer lengths, highlighting the stability of the hybrid schemes in balancing computational complexity and write endurance. These significant improvements are maintained across various pointer bit-widths, ranging from 8 to 21 bits in a pointer, as illustrated in both Figures 7 and 8 and Table III.

### C. Tree Rotations with Dynamic Reallocation

In the HART scheme, not all nodes are assigned Gray codes, which significantly mitigates computational overhead associated with address conflicts in Phase-Change Memory (PCM). By using Gray codes primarily for deeper nodes, where rotations and address conflicts are more frequent, and employing linear addressing for shallower nodes, the scheme avoids the frequent use of spare queue of addresses. This selective use of Gray codes reduces the likelihood of address conflicts and duplicates, thereby preserving computational efficiency and minimizing write overhead as compared to DFAT-Gray. The hybrid approach balances the reduced bit flips and computational simplicity, resulting in significant gains in memory endurance and performance without the severe impact on computational cost typically associated with full Gray code usage which DFAT-Gray employs. This selective strategy also explains the lower computational cost in our method.

## V. CONCLUSION

We introduce HART, a novel addressing scheme that optimizes address allocation in self-balancing tree structures by combining Depth-First Alternating Traversal (DFAT)-Gray encoding for shallower nodes with linear addressing for deeper nodes. This hybrid approach reduces computational overhead during frequent rotations at shallow tree levels, while maintaining simplicity at deeper levels. Our results show that HART reduces computational cost by up to 65%, while only slightly increasing bit flips by approximately 20% for the HART at 0.5H Threshold configuration when compared to DFAT-Gray addressing scheme. Our method reduces bit flips by 84.7% compared to random, 73.2% compared to Gray, and 76.5% compared to linear addressing. These findings not only enhance PCM performance and reliability but also highlight HART's scalability, making it a promising solution for high-performing, durable memory architectures in big data systems where efficient and reliable memory management is crucial.